\documentclass[12pt,reqno]{amsart}

\usepackage{amsmath,amsfonts,amsthm,cite,amssymb}

\usepackage{fullpage}
\usepackage[footnotesize,hang,sf]{caption}
\usepackage{times}
\newcommand{\I}{\mathrm{i}}
\newcommand{\E}{\mathrm{e}}

\usepackage{titlesec}
\titleformat{\section}
{\bfseries\scshape\centering}
{\thesection.}{.5em}{}
\titleformat{\subsection}
{\rmfamily\bfseries}
{\thesubsection}{.5em}{}
\numberwithin{equation}{section}
\newcommand{\vv}{\vec}
\usepackage{upgreek}
\newtheorem{theorem}{Theorem}
\newtheorem{prop}[theorem]{Proposition}

\begin{document}

%\special{header=psm.pro}

\renewcommand{\thefootnote}{\fnsymbol{footnote}}
%%%
\baselineskip 16pt
\parskip 8pt
\sloppy

% \noindent
% \UT

%\vspace{27pt}

%%%%%%%%%%%%%%%%%% TITLE %%%%%%%%%%%%%%

\title{Mock (False) Theta Functions as Quantum Invariants}

%%%%%%%%%%%%%%%%%%%%%%% AUTHOR(S) %%%%%%%%%%%%%%%%%%%

\author{Kazuhiro \textsc{Hikami}}

%%%%%%%%%%%%%%%%%%%%%%% ADDRESS %%%%%%%%%%%%%%%%%%%%

  \address{Department of Physics, Graduate School of Science,
%    \\
    University of Tokyo,
%    \\
    Hongo 7--3--1, Bunkyo, Tokyo 113--0033,   Japan.
    }
%     \\
    
    \urladdr{http://gogh.phys.s.u-tokyo.ac.jp/{\textasciitilde}hikami/}

    \email{\texttt{hikami@phys.s.u-tokyo.ac.jp}}

\dedicatory{
  Contribution to the Special Issue Commemorating the 200th
  Anniversary of the Birth of Carl Gustav Jacob Jacobi
}

%%%%%%%%%%%%%%%%%%%%%% DATE %%%%%%%%%%%%%%%%%%%%%%%%%%%
%(Received: \hspace{40mm})

%\vspace{18pt}
%\date{\today}
\date{June 15, 2005}

%%%%%%%%%%%%%%%%%%%%%% ABSTRACT %%%%%%%%%%%%%%%%%%%%%%
\begin{abstract}
We establish a correspondence between
the SU(2) Witten--Reshetikhin--Turaev invariant for the
Seifert manifold $M(p_1, p_2, p_3)$ and  Ramanujan's mock
theta functions.

\end{abstract}

%%%%%%%%%%%%%%%%%%%%%%% Key Words %%%%%%%%%%%%%%%%%%%%%%%%%%

%\noindent
%\textsf{Key Words:}

%%\textsl{PACS:}
%\subjclass[2000]{
%}

%%%%%%%%%%%%%%%%%%%%%%%%%%%%%%%%%%%%%%%%%%%%%%%%%%%%%%%%%
%\newpage

%%\renewcommand{\thefootnote}{\arabic{footnote}}
\maketitle
%%%%%%%%%%%%%%%%%%%%%%%%%%%%%%%%%%%%

\section{Introduction}

According to Jacobi, theory of elliptic functions was given birth in
December  23, 1751, when Euler was asked to  referee Fagnano's paper
including addition formula for the arc length of ellipse and the
lemniscate.
Since then it has been established and developed by Jacobi, Abel,
Gauss and others,
and  it now
becomes fundamental and considerable subjects in both
mathematics and physics.

One of mysterious topics of the theta functions is
Ramanujan's mock theta function (see \emph{e.g.}
Refs.~\citen{Andre79a,GEAndre89a}).
It is not modular, but it has a nice asymptotic behavior when $q$ is
root of unity.
Watson gave a proof for the third order mock theta
functions~\cite{GWats36},
though the systematic understanding and
even the meaning of ``order'' are  still missing.
In his thesis~\cite{Zweg02Thesis}, Zwegers investigated  the mock
theta functions by use of  a real analytic modular form with
half-integral weight
(see also Ref.~\citen{Zwege01a}).
He showed that the mock theta functions can be written as a sum of the
\emph{indefinite} theta functions and the Eichler integral of the
half-integral weight modular forms.
He then obtained the transformation formula for the mock theta
functions,which indicates that they have a \emph{nearly} modular property.

Nearly modular property has also revealed  in recent studies of 
the quantum
invariant~\cite{LawrZagi99a}.
Therein the SU(2)
Witten--Reshetikhin--Turaev (WRT)
invariant
$\tau_N(\mathcal{M})$~\cite{ResheTurae91a,EWitt89a}
for the Poincar{\'e} homology sphere
$\mathcal{M}=\Sigma(2,3,5)$  was identified with
a limiting value of
the Eichler integral of the modular form with half-integral weight
in
$\tau\to 1/N$,
and an exact asymptotic expansion in $N\to\infty$ was given by use of
the modular transformation.
This method is further applied to the colored Jones polynomial for torus
knot~\cite{KHikami03c} and torus link~\cite{KHikami03a},
and to the WRT invariant for the Seifert
manifold~\cite{KHikami04b,KHikami04e,KHikami05a}, and topological
invariants such as the Chern--Simons invariant, the Casson invariants,
the Reidemeister--Ray--Singer torsion,
and the Ohtsuki invariant are interpreted from the view point of the
modular form.

Purpose of this article is to show that  the mock theta functions can
be regarded as the WRT invariant
$\tau_N(\mathcal{M})$
for the Seifert manifold
$\mathcal{M}=M(p_1, p_2, p_3)$
(see, \emph{e.g.}, Ref.~\citen{JMiln75a} for the Seifert manifolds). 
Precisely the WRT invariant can be written in terms of the mock theta
functions as a limiting value  $\tau\to 1/N$ from lower half of the
complex plane.
This  coincidence was already pointed out in Ref.~\citen{LawrZagi99a}
(see also Ref.~\citen{Zwege01a})
for the case of the Poincar{\'e} homology sphere which is connected to
the fifth order
mock theta function, and
here we find  that
such relationship does exist for every order of Ramanujan's mock theta
functions.

An outline of the present article is as follows.
In Section~\ref{sec:modular} we introduce the vector-valued modular
form with weight $3/2$.
We define the Eichler integral thereof, and review a property of the
Eichler integral.
In  Sections~\ref{sec:5th} --~\ref{sec:10th},
we discuss separately Ramanujan's
mock theta functions of order 3, 5, 6, 7,
10.
Therein
we consider the mock theta functions when $q$ is outside the unit
circle, and prove that they are the Eichler integral of the
vector-valued modular form with weight $3/2$, and that a limiting
value in $\tau \to 1/N$ gives the WRT invariant for the Seifert
manifold
which was studied in Ref.~\citen{KHikami05a}.
The last section is devoted to discussions.

Hereafter we set
\begin{equation*}
  q = \E^{2 \pi \I \tau} ,
\end{equation*}
with $\tau \in \mathbb{H}$, and
we use a standard notation of $q$-calculus;
\begin{gather*}
  (x)_n = (x ; q)_n =
  \prod_{k=1}^n \left(1 - x \, q^{k-1} \right) ,
  \\[2mm]
  (a, b, \cdots ; q)_n
  =
  (a ; q)_n \, (b ; q)_n \cdots ,
  \\[2mm]
  \begin{bmatrix}
    n \\
    m
  \end{bmatrix}_q
  =
  \frac{(q)_n}{(q)_{n-m} \, (q)_m} .
\end{gather*}
For our convention, we list some identities (see, \emph{e.g.},
Ref.~\citen{Andre76});
\begin{itemize}
\item $q$-binomial theorem,
  \begin{equation}
    \label{Gauss_polynomial}
    (-z)_N =
    \sum_{m=0}^N q^{m(m-1)/2} \,
    \begin{bmatrix}
      N \\
      m
    \end{bmatrix}_q \,
    z^m ,
  \end{equation}

\item $q$-binomial series,
  \begin{equation}
    \label{q-binomial_over}
    \frac{1}{(z)_N}
    =
    \sum_{m=0}^\infty
    \begin{bmatrix}
      N+m-1 \\
      m
    \end{bmatrix}_q \,
    z^m ,
  \end{equation}

\item $q$-binomial formula,
  \begin{equation}
    \label{q-binomial}
    \sum_{n=0}^\infty
    \frac{
      (a)_n
    }{
      (q)_n
    } \,
    z^n
    =
    \frac{( a \, z)_\infty}{(z)_\infty} ,
  \end{equation}
\item the  Euler identity,
  which follows from ~\eqref{Gauss_polynomial} in $N\to\infty$,
  \begin{equation}
    \label{Euler_identity}
    \sum_{m=0}^\infty
    \frac{q^{m(m-1)/2}}{(q)_m} \,
    z^m
    =
    ( - z)_\infty ,
  \end{equation}

\item the  Jacobi triple product identity,
  \begin{equation}
    \label{triple_product}
    \sum_{k \in \mathbb{Z}}
    (-1)^k \, q^{k^2/2} \, z^k
    =
    (q, z^{-1} \, q^{1/2} , z \, q^{1/2} ; q)_\infty .
  \end{equation}
\end{itemize}
%%%%%%%%%%%%%%%%%%%
\section{Modular Form}
\label{sec:modular}

We define an  odd periodic function $\psi_{2P}^{(a)}(n)$
with modulus $ 2\, P$
\begin{equation}
  \label{periodic_psi}
  \psi_{2 P}^{(a)}(n)
  =
  \begin{cases}
    \pm 1  ,
    &
    \text{for $n \equiv \pm a \mod 2 \, P$,}
    \\[3mm]
    0 ,
    &
    \text{otherwise,}
  \end{cases}
\end{equation}
for $a \in \mathbb{Z}$.
When we introduce the function 
\begin{equation}
  \label{define_modular_Psi}
  \Psi_P^{(a)}(\tau)
  =
  \frac{1}{2} \,
  \sum_{n \in \mathbb{Z}}
  n \, \psi_{2 P}^{(a)}(n) \,
  q^{\frac{n^2}{4 P}} ,
\end{equation}
this becomes a vector-valued modular form with weight $3/2$
satisfying
\begin{gather}
  \label{Psi_under_S}
  \Psi_P^{(a)}(\tau)
  =
  \left(
    \frac{\I}{\tau}
  \right)^{3/2} \,
  \sum_{b=1}^{P-1} \mathbf{M}(P)_a^b \,
  \Psi_P^{(b)}(-1/\tau) ,
  \\[2mm]
  \label{Psi_under_T}
  \Psi_P^{(a)}(\tau+1)
  =
  \E^{\frac{a^2}{2 P} \pi \I} \,
  \Psi_P^{(a)}(\tau) ,
\end{gather}
where $\mathbf{M}(P)$ is  a $(P-1)\times(P-1)$ matrix
\begin{equation}
  \mathbf{M}(P)_a^b
  =
  \sqrt{\frac{2}{P}} \,
  \sin
  \left(
    \frac{a \, b}{P} \, \pi
  \right) .
\end{equation}

The Eichler integral  of this  set of the modular forms with
half-integral weight is then defined
as a half-integration of $\boldsymbol{\Psi}_{P}^{(a)}(\tau)$ with
respect to $\tau$ by~\cite{LawrZagi99a,KHikami03a}
\begin{equation}
  \label{define_Eichler}
  \widetilde{\Psi}_P^{(a)}(\tau)
  =
  \sum_{n=0}^\infty
  \psi_{2 P}^{(a)}(n) \,
  q^{\frac{n^2}{4 P}} .
\end{equation}
The Eichler integral of the \emph{integral}-weight modular form is known to
have a nearly modular property (see, \emph{e.g.},
Ref.~\citen{SLang76Book}).
To see this nearly modular property in our case,
we introduce another Eichler integral following Ref.~\citen{LawrZagi99a};
\begin{equation}
  \label{hat_Eichler}
  \widehat{\Psi}_P^{(a)}(z)
  =
  \frac{1}{
    \sqrt{ 2 \, P \, \I}
  } \,
  \int_{\overline{z}}^{\I \infty}
  \frac{
    \Psi_P^{(a)}(\tau)
  }{
    \sqrt{\tau - z}
  } \,
  \mathrm{d} \tau ,
\end{equation}
which is defined for $z$ in the lower half plane, $z\in\mathbb{H}^-$.
We see that the modular property  of
$\Psi_P^{(a)}(\tau)$, especially the modular
${S}$-transformation~\eqref{Psi_under_S},
leads
a nearly modular property of the Eichler integral as~\cite{LawrZagi99a}
(see also Refs.~\citen{KHikami03a,Zweg02Thesis})
\begin{gather}
  \label{modular_Psi_hat}
  \widehat{\Psi}_P^{(a)}(z)
  +
  \frac{  \  1 \   }{\sqrt{ \I \, z}} \,
  \sum_{b=1}^{P-1}
  \mathbf{M}(P)_b^a \, \widehat{\Psi}_P^{(b)}
  (- 1/z)
  =
%  r_{\Psi_P}^{(a)}(z; 0)
  \frac{1}{\sqrt{ 2 \, P \, \I}}
  \,
  \int_{0}^{\I \infty}
  \frac{
    \Psi_P^{(a)}(\tau)
  }{
    \sqrt{\tau - z}
  } \,
  \mathrm{d} \tau .
\end{gather}
% where
% \begin{gather*}
%   r_{\Psi_P}^{(a)}(z; \alpha)
%   =
%   \frac{1}{\sqrt{ 2 \, P \, \I}}
%   \,
%   \int_{\alpha}^\infty
%   \frac{
%     \Psi_P^{(a)}(\tau)
%   }{
%     \sqrt{\tau - z}
%   } \,
%   \mathrm{d} \tau
% \end{gather*}
%  with $\alpha\in\mathbb{Q}$.
Then taking a limit $\tau, z \to 1/N$  for $N\in \mathbb{Z}$
in the Eichler integrals $\widetilde{\Psi}_P^{(a)}(\tau)$
and $\widehat{\Psi}_P^{(a)}(z)$, we
find both Eichler integrals coincide in this limit, and we obtain
an asymptotic expansion in $N\to\infty$ as
\begin{equation}
  \label{nearly_modular_Psi_tilde}
  \widetilde{\Psi}_P^{(a)}(1/N)
  +
  \sqrt{\frac{N}{\I}} \,
  \sum_{b=1}^{P-1}
  \mathbf{M}(P)_a^b \,
  \widetilde{\Psi}_P^{(b)}(-N)
  \simeq
  \sum_{k=0}^\infty
  \frac{L( -2 \, k , \psi_{2 P}^{(a)})}{k!} \,
  \left(
    \frac{\pi \, \I}{2 \, P \, N}
  \right)^k ,
\end{equation}
where $L(k,\psi_{2P}^{(a)})$ denotes the Dirichlet $L$-function
associated with $\psi_{2 P}^{(a)}(n)$
defined in  \eqref{periodic_psi}.
We note that for $N\in \mathbb{Z}$ and $1 \leq a \leq P-1$
we have
\begin{gather}
  \label{Eichler_at_over_N}
  \widetilde{\Psi}_P^{(a)}(1/N)
  =
  -\sum_{k=0}^{2 P N}
  \psi_{2 P}^{(a)}(k) \,
  \E^{\frac{k^2}{2 P N} \pi \I} \,
  B_1
  \left(
    \frac{k}{2 \, P \, N}
  \right) ,
  \\[2mm]
  \widetilde{\Psi}_P^{(a)}(N)
  =
  \left(
    1- \frac{a}{P}
  \right) \,
  \E^{\frac{a^2}{2 P} \pi \I N} ,
\end{gather}
where the Bernoulli polynomial is $B_1(x)=x-\frac{1}{2}$.
We note that the Eichler integral
$\widetilde{\Psi}_P^{(P-1)}(1/N)$ is the spefic value of the
$N$-colored Jones polynomial for torus link
$\mathcal{T}_{2,2 P}$~\cite{KHikami03a}.

The nearly modular property~\eqref{modular_Psi_hat} resembles with the
transformation properties of the mock theta functions~\cite{GWats36,Zweg02Thesis}.
In fact the right hand side of~\eqref{modular_Psi_hat} can be
rewritten in terms of the Mordell integrals~\cite{Zweg02Thesis},
and  it suggests a connection between the Eichler integrals of the
weight $3/2$ modular form and the
mock theta functions.
%%%%%%%%%%%%%%%%%%
%%%%%%%%%%%%
\section{The 5th Order Mock Theta Functions}
\label{sec:5th}

We start from the fifth order mock theta functions~\cite{GEAndre86c}.
Ramanujan defined 10 functions, and here we treat 2 functions among
them defined by
\begin{gather}
  \chi_0(q)
  =
  \sum_{n=0}^\infty
  \frac{q^n}{(q^{n+1})_n} ,
  \label{define_chi0}
  \\[2mm]
  \chi_1(q)
  =
  \sum_{n=0}^\infty
  \frac{q^n}{(q^{n+1})_{n+1}} ,
  \label{define_chi1}
\end{gather}
Other 8 functions can be given by use of these functions and
theta functions~\cite{GWats37,Hicker88a}.

% When we set
% \begin{equation}
%   \boldsymbol{\Phi}_5(\tau)
%   =
%   \begin{pmatrix}
%     q^{-\frac{1}{120}} \,
%     \left(
%       \chi_0(q) - 2
%     \right)
%     \\[2mm]
% %
%     q^{\frac{71}{120}} \,
%     \chi_1(q)
%   \end{pmatrix}
% \end{equation}
% the transformation properties of these functions are given
% by~\cite{GordMcIn03a}
% \begin{multline}
%   \boldsymbol{\Phi}_5(\tau)
%   =
%   -\sqrt{\frac{\I}{\tau}} \,
%   \frac{2}{\sqrt{5}}
%   \begin{pmatrix}
%     \sin \left( \frac{\pi}{5} \right) &
%     \sin \left( \frac{2 \pi}{5} \right)
%     \\[2mm]
%     \sin \left( \frac{2 \pi}{5} \right) &
%     - \sin \left( \frac{ \pi}{5} \right)
%   \end{pmatrix}
%   \cdot
%   \boldsymbol{\Phi}_5(-1/\tau)
%   \\
%   -
%   \sqrt{\frac{\tau}{\I}} \,
%   2 \sqrt{135}
%   \int_0^\infty
%   \E^{15 \pi \I \tau x^2} \,
%   \begin{pmatrix}
%     \frac{
%       \cos(9 \pi \tau x) \, \cos(5 \pi \tau x)
%     }{
%       \cos(15 \pi \tau x)
%     }
%     \\[2mm]
%     \frac{
%       \cos(3 \pi \tau x) \, \cos(5 \pi \tau x)
%     }{
%       \cos(15 \pi \tau x)
%     }
%   \end{pmatrix}
%   \mathrm{d} x
% \end{multline}

As was noticed  in Ref.~\citen{LawrZagi99a},
definitions~\eqref{define_chi0} and~\eqref{define_chi1}
can be extended to outside the unit
circle
$|q|>1$, and we define
new functions by substituting $1/q$ in place of $q$;
\begin{gather}
  \label{chi0_inverse}
  \chi_0^*(q)
  = 2-\chi_0(1/q)=
  2- 
  \sum_{n=0}^\infty
  (-1)^n
  \frac{q^{\frac{3}{2} n^2 - \frac{1}{2} n}}{(q^{n+1})_n} ,
  \\[2mm]
  \label{chi1_inverse}
  \chi_1^*(q)
  =
  q^{-1} \, \chi_1(1/q)
  =
  \sum_{n=0}^\infty
  (-1)^n 
  \frac{q^{\frac{3}{2} n (n+1)}}{
    (q^{n+1})_{n+1}
  } .
\end{gather}

\begin{prop}
  \begin{gather}
    \label{chi0_Eichler}
    \chi_0^*(q)
    =
    \sum_{n=0}^\infty \chi_{60}^{(1,1,1)}(n) \, q^{\frac{n^2-1}{120}} ,
    \\[2mm]
    \label{chi1_Eichler}
    \chi_1^*(q)
    =
    \sum_{n=0}^\infty \chi_{60}^{(1,1,2)}(n) \, q^{\frac{n^2-49}{120}}
    ,
  \end{gather}
  where $\chi_{60}^{\vv{\ell}}(n)$ is an odd periodic function with
  modulus $60$ defined by
  \begin{gather*}
    \chi_{60}^{(1,1,1)}(n)
    =
    \psi_{60}^{(1)}(n) +     \psi_{60}^{(11)}(n)
    +    \psi_{60}^{(19)}(n) +     \psi_{60}^{(29)}(n) ,
    \\[2mm]
    \chi_{60}^{(1,1,2)}(n)
    =
    \psi_{60}^{(7)}(n) +     \psi_{60}^{(13)}(n)
    +    \psi_{60}^{(17)}(n) +     \psi_{60}^{(23)}(n) .
  \end{gather*}
\end{prop}

\begin{proof}
  These can be proved by applying the Baily chain~\cite{Baile49a}
  (see also  Ref.~\citen{Andre84a});
  if for $n \geq 0$
  the Bailey pair $(\alpha_n, \beta_n)$ satisfies
  \begin{equation}
    \label{Bailey_pair}
    \sum_{k=0}^n
    \frac{\alpha_k}{(q)_{n-k} \, (x \, q)_{n+k}}
    =
    \beta_n  ,
  \end{equation}
  we have
  \begin{multline}
    \sum_{n=0}^\infty
    \frac{
      (\rho_1)_n \, (\rho_2)_n
    }{
      \left( \frac{x \, q}{\rho_1} \right)_n \,
      \left( \frac{x \, q}{\rho_2} \right)_n 
    } \,
    \left(
      \frac{x \, q}{\rho_1 \, \rho_2}
    \right)^n \, \alpha_n \,
    \frac{1}{
      (q)_{N-n} \, (x \, q)_{N+n}
    }
    \\
    =
    \sum_{n=0}^\infty
    \frac{
      (\rho_1)_n \, (\rho_2)_n \,
      \left( \frac{x \, q}{\rho_1 \, \rho_2} \right)_{N-n}
    }{
      (q)_{N-n}
    } 
    \,
    \left(
      \frac{
        x \, q
      }{
        \rho_1 \, \rho_2
      }
    \right)^n \,
    \beta_n \,
    \frac{1}{
      \left( \frac{x \, q}{\rho_1} \right)_N
      \,
      \left( \frac{x \, q}{\rho_2} \right)_N
    } ,
  \end{multline}
  which reduces to
  \begin{equation}
    \label{simple_Bailey}
    (1-x) \,
    \sum_{n=0}^\infty
    \frac{(q)_n}{(x)_n} \, x^n \, \alpha_n \,
    (-1)^n \, q^{\frac{1}{2} n (n-1)}
    =
    \sum_{n=0}^\infty
    (q)_n \, x^n \, \beta_n \,
    (-1)^n \, q^{\frac{1}{2} n (n-1)} ,
  \end{equation}
  by setting $\rho_1=q$, $\rho_2\to\infty$ and $N\to\infty$.

  Eqs.~\eqref{chi0_Eichler} and~\eqref{chi1_Eichler} follow
  immediately when
  we use the Bailey pairs,
  A(5) and A(8) in  Slater's list~\cite{LJSlater51}.
  See also Ref.~\citen{LJRogers17a},
  where these functions are called the \emph{false} theta function.
% %%%%%%%%%%%%
%    As an example of the Bailey pair satisfying  \eqref{Bailey_pair}
%    with $x=1$, we have
%    \begin{align}
%      \beta_n
%      & =
%      \frac{q^{n^2}}{(q)_{2 n}}
%      &
%      \begin{cases}
%        \alpha_{3k-1 > 0} = - q^{k (3k-1)}
%        \\[2mm]
%        \alpha_{3k >0} =  q^{k (3k-1)} +  q^{k (3k+1)} ,
%        \qquad
%        \alpha_0 = 1
%        \\[2mm]
%        \alpha_{3k+1  >0 } = - q^{k (3k+1)}
%      \end{cases}
%    \end{align}
%    which is  A(5) in Slater's list~\cite{LJSlater51}.
%    Substituting  \eqref{simple_Bailey}, we obtain
%     \eqref{chi0_Eichler}.
%    To prove  \eqref{chi1_Eichler}, we only need to use
%    A(8) in Slater's list
%    with $x=q$;
%    \begin{align}
%      \beta_n
%      & =
%      \frac{q^{n(n+1)}}{(q^2)_{2 n}}
%      &
%      \begin{cases}
%        \alpha_{3k-1 >0} = - q^{k (3k-2)}
%        \\[2mm]
%        \alpha_{3k  \geq 0} =    q^{k (3k+2)}
%        \\[2mm]
%        \alpha_{3k+1  > 0} = - q^{(k+1) (3k+1)} -  q^{k (3k+2)}
%      \end{cases}
%    \end{align}
%    Substitution for  \eqref{chi1_inverse} proves
%     \eqref{chi1_Eichler}.
\end{proof}

According to Ref.~\citen{LawrZagi99a},
these \emph{false} theta functions 
should rather be identified as
the Eichler integral.
Namely when we set
\begin{equation*}
  \widetilde{\boldsymbol{\Phi}}_{2,3,5}(\tau)
  =
  \begin{pmatrix}
    q^{\frac{1}{120}} \, \chi_0^*(q)
    \\[2mm]
    q^{\frac{49}{120}} \, \chi_1^*(q)
  \end{pmatrix} ,
\end{equation*}
which has a form like  \eqref{define_Eichler} due to
~\eqref{chi0_Eichler} --~\eqref{chi1_Eichler},
it can be regarded as the Eichler integral
of the vector-valued
modular form with weight $3/2$;
\begin{equation}
  \boldsymbol{\Phi}_{2,3,5}(\tau)
  =
  \frac{1}{2} \sum_{n\in\mathbb{Z}}
  n \, 
  \begin{pmatrix}
    \chi_{60}^{(1,1,1)}(n)
    \\[2mm]
    \chi_{60}^{(1,1,2)}(n) 
  \end{pmatrix}
  \,
  q^{\frac{n^2}{120}} .
\end{equation}
The  modular $S$- and $T$-matrices
under $\tau \to -1/\tau$ and $\tau \to \tau+1$
are respectively given by
\begin{align}
  \label{S_T_235}
    \mathbf{S}
    & =
    \frac{2}{\sqrt{5}}\,
    \begin{pmatrix}
      \sin \left( \frac{\pi}{5} \right)
      &       \sin \left( \frac{2 \, \pi}{5} \right) 
      \\[2mm]
      \sin \left( \frac{2 \, \pi}{5} \right) &
      -   \sin \left( \frac{\pi}{5} \right)
    \end{pmatrix}
    ,
    & 
    \mathbf{T}
    & =
    \begin{pmatrix}
      \E^{ \frac{1}{60} \pi \I} &
      \\[2mm]
      &
      \E^{ \frac{49}{60} \pi \I}
    \end{pmatrix} .
  \end{align}
This shows that the modular form 
$[ \eta(\tau) ]^{-1/5} \cdot \boldsymbol{\Phi}_{2,3,5}(\tau)$ with
rational weight
$7/5$ is on the principal  congruence subgroup $\Gamma(5)$.

Intriguing is that
the Eichler integral $\widetilde{\boldsymbol{\Phi}}_{2,3,5}(\tau)$
gives the WRT invariant for the Poincar{\'e}
homology sphere in
a limit of $q$ being the $N$-th root of unity
as was  proved by Lawrence and Zagier~\cite{LawrZagi99a}.

\begin{theorem}
  \begin{equation}
    \label{235_mock}
    \E^{\frac{2 \pi \I}{N}} \,
    \left(
      \E^{\frac{2 \pi \I}{N}} - 1
    \right) \,
    \tau_N
    \left( \Sigma(2,3,5) \right)
    =
    1 - \frac{1}{2} \, \chi_0^*(\E^{\frac{2 \pi \I}{N}}) .
  \end{equation}
\end{theorem}

Here the WRT invariant $\tau_N(\mathcal{M})$ for 3-manifold
$\mathcal{M}$ is normalized to be
\begin{gather*}
  \tau_N(S^3)
  =1 ,
  \\[2mm]
  \tau_N(S^2 \times S^1)
  =
  \sqrt{ \frac{N}{2}} \,
  \frac{1}{
    \sin\left( \pi /N \right)
  } .
\end{gather*}
We should remark that we have
the icosahedral symmetry
of  the group
$\Gamma(5)$ and the fundamental group of the
Poincar{\'e} sphere.
As a consequence of~\eqref{235_mock}, we see that the Eichler integral in $\tau\to 1/N$
has a nearly modular property~\eqref{nearly_modular_Psi_tilde}
replacing $\mathbf{M}(P)$ and $\psi_{2P}^{(a)}(n)$ with
$\mathbf{S}$~\eqref{S_T_235} and $\chi_{60}^{\vv{\ell}}(n)$
respectively, and $P=30$.

We
can obtain  an explicit form of the
quantum invariant~\eqref{235_mock} as a linear combination of
 \eqref{Eichler_at_over_N}
by taking a limiting value of the Eichler integral
$\widetilde{\boldsymbol{\Phi}}_{2,3,5}(\tau)$
in $\tau \to 1/N$~\cite{LawrZagi99a}.
To derive this  invariant in terms of  $q$-series,
it is useful to rewrite 
the Eichler integral in the form such that the infinite sum terminates
at  the finite sum in the case of $q$ being  root of unity.

\begin{prop}
  \begin{gather}
    \label{Le_expression}
    \chi_0^*(q)
    =
    1 + q \, \sum_{m=0}^\infty q^{2 m} \, (q^{m+1})_m
    =
    \sum_{n=0}^\infty q^n \, (q^n)_n ,
    \\[2mm]
    \chi_1^*(q)
    =
    \sum_{n=0}^\infty q^n \, (q^{n+1})_n .
  \end{gather}
\end{prop}
\begin{proof}
   {}From  \eqref{chi0_inverse} we compute as follows;
   \begin{align*}
     \chi_0^*(q)
     & =
     1+  \sum_{n=0}^\infty (-1)^n \frac{
       q^{\frac{1}{2} (n+1)(3n+2)}
     }{
       (q^{n+2})_{n+1}
     }
     \\
     & =
     1+
     q \sum_{n=0}^\infty
     \frac{(q^{2n+3})_\infty}{(q^{n+2})_\infty} \,
     (-1)^n \,
     q^{\frac{3}{2} n^2 + \frac{5}{2} n}
     \\
     & =
     % \stackrel{\eqref{q-binomial}}{=}
     1 + q
     \sum_{n,k=0}^\infty
     (-1)^n \,  q^{\frac{3}{2} n^2 + \frac{5}{2} n}
     \frac{(q^{n+1})_k}{(q)_k} \,
     q^{(n+2)k}
     \tag{by \eqref{q-binomial}}
     \displaybreak[0]
     \\
     & =
     1 + q \sum_{m \geq n\geq 0}^\infty
     q^{2m} \,
     \begin{bmatrix}
       m       \\
       n
     \end{bmatrix}_q
     \,
     (-q^{m+1})^n \, q^{\frac{1}{2} n(n-1)}
     \\
     & =
%     \stackrel{\eqref{Gauss_polynomial}}{=}
     1 + q \sum_{m=0}^\infty
     q^{2m } \, (q^{m+1})_m
     \tag{by \eqref{Gauss_polynomial}}
     \displaybreak[0]
     \\
     & =
     1 +
     q \, (q)_\infty \sum_{m=0}^\infty
     \frac{q^{2m}}{(q)_m} \,
     \frac{1}{(q^{2m+1})_\infty}
     \\
     & =
%     \stackrel{\eqref{q-binomial}}{=}
     1+ q \, (q)_\infty
     \sum_{n,m=0}^\infty \frac{q^n}{(q)_n} \, \frac{q^{2nm + 2
         m}}{(q)_m}
     \tag{by \eqref{q-binomial}}
     \displaybreak[0]
     \\
     & =
%     \stackrel{\eqref{q-binomial}}{=}
     1 + q \sum_{n=0}^\infty q^n \, (q^{n+1})_{n+1} .
     \tag{by \eqref{q-binomial}}
   \end{align*}
   Identity~\eqref{chi1_inverse} can be proved in the same manner.
 \end{proof}

We see that,
by definition of the $q$-product,
the infinite sum in the
expression~\eqref{Le_expression}  terminates at the finite order when
$q$ is  root of unity.
Furthermore the expression~\eqref{Le_expression}
exactly coincides with  the form given by Le in
Refs.~\citen{TQLe03a,TQLe05unpublish},
where the WRT invariant was computed by use of Habiro's cyclotomic
expansion of the $N$-colored Jones polynomial for
trefoil~\cite{Habiro00a,KHabi02a,GMasb03a,LawrDRon05a}
\begin{equation}
  \label{cyclotomic_Jones}
  J_{\mathcal{T}_{2,3}}(N)
  =
  \sum_{k=0}^\infty
  q^{-k(k+2)} \,
  (q^{1-N})_k \, (q^{1+N})_k ,
\end{equation}
where we have used the normalized colored Jones polynomial s.t.
$J_{\text{unknot}}(N)=1$.
It is known  to be
rewritten as~\cite{KHabi02a,TQLe03a,KHikami04a}
\begin{equation}
  \label{our_Jones}
  J_{\mathcal{T}_{2,3}}(N)
  =
  q^{1-N} \,
  \sum_{k=0}^\infty
  q^{-k N} \,
  (q^{1-N})_k ,
\end{equation}
{}from which,
simply applying $(+1)$-surgery on this expression following
Ref.~\citen{ResheTurae91a}
(see also Ref.~\citen{TQLe05unpublish}),
we obtain another expression 
\begin{equation}
  \chi_0^*(q)
  =
  1+q \sum_{k \geq n \geq 0}^\infty
  (-1)^n \,
  \begin{bmatrix}
    k \\
    n
  \end{bmatrix}_q
  \,
  q^{k(k+1) + \frac{1}{2}n (3n+5)+k n} .
\end{equation}
This expression
also reduces to finite sum when $q$ is root of unity.
We do not have a direct proof of this $q$-series identity at  present.

To close this section we comment on the $q$-hypergeometric 
type generating function of the $L$-function at negative values.
Studies on such generating functions have been developed since
Ref.~\citen{DZagie01a}.
See Refs.~\citen{LoveKOno02a,KOno04Book,AndrUrroOnok01a,KHikami02c,KHikami02b}.
Based on  \eqref{Le_expression}, we have the following as a power
series in $t$;
   \begin{gather}
     \begin{aligned}
       \E^{-t/120}
       \sum_{n=0}^\infty
       \E^{-n t} \,
       (1 - \E^{-n t}) \,
       (1 - \E^{-(n+1) t})
       & \cdots
       (1 - \E^{-(2n-1) t})
       \\
       &
       =
       \frac{1}{2}
       \sum_{k=0}^\infty
       \frac{
         L\left(-2 \, k , \chi_{60}^{(1,1,1)}\right)
       }{
         k!
       } \,
       \left(
         \frac{ -t}{120}
       \right)^k ,
     \end{aligned}
     \\[2mm]
     \begin{aligned}
       \E^{-49 t/120}
       \sum_{n=0}^\infty
       \E^{-n t} \,
       (1 - \E^{-(n+1) t}) \,
       (1 - \E^{-(n+2) t})
       & \cdots
       (1 - \E^{-2 n t})
       \\
       &    =
       \frac{1}{2}
       \sum_{k=0}^\infty
       \frac{
         L\left(-2 \, k , \chi_{60}^{(1,1,2)}\right)
       }{
         k!
       } \,
       \left(
         \frac{ -t}{120}
       \right)^k ,
     \end{aligned}
   \end{gather}
   where
   \begin{gather*}
     2     \frac{\cos(5 \, x) \, \cos(9 \, x)
     }{
       \cos(15 \, x)
     }
     =
     \sum_{k=0}^\infty
     \frac{L
       \left( -2 \, k, \chi_{60}^{(1,1,1)}
       \right)
     }{
       ( 2 \, k  )!
     } \,
     (-1)^k \, x^{2 k} ,
     \\[2mm]
     2 \,\frac{\cos(5 \, x) \, \cos(3 \, x)
     }{
       \cos(15 \, x)
     }
     =
     \sum_{k=0}^\infty
     \frac{L
       \left( -2 \, k, \chi_{60}^{(1,1,2)}
       \right)
     }{
       ( 2 \, k  )!
     } \,
     (-1)^k \, x^{2 k}  .
   \end{gather*}

%%%%%%%%%%%%%%%%%%%
\section{The 3rd Order Mock Theta Functions}
\label{sec:3rd}

We next study the third order mock theta functions.
Among 4  Ramanujan's mock  theta functions of the third order,
we study  functions defined by
\begin{gather}
  \phi(q) 
  =
  \sum_{n=0}^\infty
  \frac{q^{n^2}}{
    \left(
      -q^2; q^2
    \right)_n} ,
  \\[2mm]
  \nu(q)
  =
  \sum_{n=0}^\infty
  \frac{q^{n(n+1)}}{
    \left(
      -q ; q^2
    \right)_{n+1}
  } .
\end{gather}
Watson gave 3 more functions in Ref.~\citen{GWats36},
which were in ``lost'' notebook~\cite{Ramanujan87Book}.
Other third order mock theta functions can be given from these
functions and the theta functions
as was proved in Ref.~\citen{GWats36}.

Both defining $q$-series
$\phi(q)$ and $\nu(q)$
converge not only inside, but also outside the unit
circle as was noticed in Ref.~\citen{Zwege01a}, and we define new
functions replacing $q$ by
$1/q$ as
\begin{gather}
  \label{3_phi_star}
  \phi^*(q)
 = \phi(1/q)
  =
  \sum_{n=0}^\infty
  \frac{q^n}{(-q^2 ; q^2)_n} ,
  \\[2mm]
  \nu^*(q)
  =  q^{-1} \, \nu(1/q)
  =
  \sum_{n=0}^\infty
  \frac{
    q^{n}
  }{
    (-q ; q^2)_{n+1}
  } .
\end{gather}
As in the case of the 5th order mock theta functions, 
these are the \emph{false} theta functions in a sense of
Rogers
as was proved in Ref.~\citen{Andre79a}.
\begin{prop}
  \begin{gather}
    \label{phi_Eichler}
    \phi^*(q)
    =
    \sum_{n=0}^\infty
    \chi_{24}^{(1)}(n) \,
    q^{\frac{1}{24} ( n^2 - 1)} ,
    \\[2mm]
    \label{nu_Eichler}
    \nu^*(q)
    =
    \sum_{n=0}^\infty
    \chi_{24}^{(2)}(n) \,
    q^{\frac{1}{24} ( n^2 - 16)} ,
    \\[2mm]
    \label{phi_Eichler_minus}
    \phi^*(-q)
    =
    \sum_{n=0}^\infty
    \psi_{6}^{(1)}(n) \,
%  \chi_{24}^{(3)}(n) \,
    q^{\frac{1}{24} ( n^2 - 1)} ,
\end{gather}
where $\chi_{24}^{(a)}(n)$ is an odd periodic function with modulus
$24$ defined by
\begin{equation*}
  \begin{aligned}
    \chi_{24}^{(1)}(n)
    & =
    \psi_{24}^{(1)}(n) +     \psi_{24}^{(5)}(n)
    +     \psi_{24}^{(7)}(n) +     \psi_{24}^{(11)}(n) ,
    \\[2mm]
    \chi_{24}^{(2)}(n)
    & =
    \psi_{24}^{(4)}(n) +     \psi_{24}^{(8)}(n) ,
%     \\[2mm]
% %
%     \chi_{24}^{(3)}(n)
%     & =
%     \psi_{24}^{(1)}(n) -     \psi_{24}^{(5)}(n)
%     +     \psi_{24}^{(7)}(n) -     \psi_{24}^{(11)}(n)
  \end{aligned}
\end{equation*}
and $\psi_{2P}^{(a)}(n)$ is defined in~\eqref{periodic_psi}.
\end{prop}

Such \emph{false} theta  functions can be  defined based on some of
other mock theta
functions of the third order.
We recall the definitions~\cite{GWats36,Ramanujan87Book};
\begin{gather}
  f(q)
  =
  \sum_{n=0}^\infty
  \frac{q^{n^2}}{
    [ (-q)_n ]^2
  } ,
  \\[2mm]
  \label{define_omega}
  \omega(q)
  =
  \sum_{n=0}^\infty
  \frac{
    q^{2 n (n+1)}
  }{
    \left[
      (q ; q^2)_{n+1}
    \right]^2
  } ,
  \\[2mm]
  \upchi(q)
  =
  \sum_{n=0}^\infty
  \frac{
    q^{n^2}
  }{
    (1 -q +q^2) \, (1-q^2+q^4) \cdots
    (1-q^n + q^{2n})
  } ,
  \\[2mm]
  \varrho(q)
  =
  \sum_{n=0}^\infty
  \frac{
    q^{2n(n+1)}
  }{
    (1+q+q^2) (1+q^3+q^6) \cdots
    (1+q^{2n+1} +q^{4n+2})
  } .
\end{gather}
As was proved in Ref.~\citen{NJFine88Book}, we have
\begin{gather}
  f(q)
  =
  2- \sum_{n=0}^\infty (-1)^n \frac{q^n}{(-q)_n} ,
  \\[2mm]
  \omega(q)
  =
  \sum_{n=0}^\infty
  \frac{q^n}{(q;q^2)_{n+1}} ,
  \\[2mm]
  \upchi(q)
  =
  1 - \E^{2 \pi \I/3}
  \sum_{n=1}^\infty
  \frac{\E^{\frac{\pi \I}{3} n} \, q^n}{
    (- \E^{2 \pi \I/3} \, q)_n
  } ,
  \\[2mm]
  \varrho(q)
  =
  \sum_{n=0}^\infty
  \frac{
    \E^{- \frac{2}{3} \pi \I n} \, q^n
  }{
    (\E^{2 \pi \I/3} \, q ; q^2)_{n+1}
  } .
\end{gather}
We can  extend these defining $q$-series into $|q|>1$, and define new
functions;
\begin{gather}
  f^*(q) =
  f(1/q)
  =
  2 - \sum_{n=0}^\infty
  (-1)^n \,
  \frac{
    q^{n(n-1)/2}
  }{
    (-q)_n
  } ,
  \\[2mm]
  \omega^*(q) = - q^{-1} \, \omega(1/q)
  =
    \sum_{n=0}^\infty
  (-1)^n \,
  \frac{q^{n(n+1)}}{
    (q; q^2)_{n+1}
  } ,
  \\[2mm]
  \upchi^*(q)
  =
  \upchi(1/q)
  =
  1 - \E^{2 \pi \I/3}
  \sum_{n=1}^\infty
  \frac{
    \E^{- \frac{\pi \I}{3} n} \,
    q^{n(n-1)/2}
  }{
    ( \E^{\frac{\pi \I}{3}} \, q)_n
  } ,
  \\[2mm]
  \varrho^*(q)
  = \varrho(1/q)
  =
  \sum_{n=0}^\infty
  \frac{
    \E^{-\frac{\pi \I}{3} n} \,
    q^{n(n+1)}
  }{
    (\E^{-2 \pi \I/3} \, q; q^2)_{n+1}
  } .
\end{gather}

By applying the Bailey chain method with 
the pairs such as  C(7) and C(5) in
Slater's
list~\cite{LJSlater51},
% or Rogers's identity~\cite{LJRogers17a},
% \begin{equation}
%   1 + \sum_{n=1}^\infty
%   (-1)^n
%   \frac{q^{n (n+1)/2}}{
%     (z \, q)_n
%   } \,
%   z^{2 n}
%   =
%   \sum_{k=0}^\infty
%   (-1)^k \, z^{3 k} \, q^{k(3k+1)/2} \,
%   \left(
%     1 - z^2 \, q^{2 k+1}
%   \right)
% \end{equation}
we obtain~\cite{LJRogers17a} the following;
\begin{prop}
  \begin{gather}
    f^*(q)
    =
    2 \sum_{n=0}^\infty
    \psi_6^{(1)}(n) \, q^{\frac{1}{24} (n^2-1)} ,
    \\[2mm]
    \label{omega_star_Eichler}
    \omega^*(q)
    =
    \sum_{n=0}^\infty
    \left(
      \psi_{6}^{(1)}(n)
      +
      \psi_6^{(2)}(n)
    \right) \,
    q^{\frac{1}{3}(n^2 - 1)} ,
    \\[2mm]
    \upchi^*(q)
    =
    \sum_{n=0}^\infty
    \psi_6^{(1)}(n) \, q^{\frac{1}{24} (n^2 - 1)} \,
    \left(
      1+ \E^{- \frac{2}{3} \pi \I n}
    \right)  ,
    \\[2mm]
    \varrho^*(q)
    =
    \sum_{n=0}^\infty
    \left(
      \psi_6^{(1)}(n) +       \psi_6^{(2)}(n)
    \right) \,
    q^{\frac{1}{3} ( n^2 - 1)} \,
    \E^{\frac{2}{3} \pi \I (1-n)} .
  \end{gather}
\end{prop}

We remark that,
comparing with~\eqref{nu_Eichler}, we find
\begin{equation}
  \omega^*(q^2) = \nu^*(q)  .
\end{equation}

To study the transformation property of these functions using
 \eqref{nearly_modular_Psi_tilde}, it is better
to regard these functions as the Eichler integrals.
We define
\begin{equation}
  \label{Eichler_234}
  \widetilde{\boldsymbol{\Phi}}_{2,3,4}(\tau)
  =
  \begin{pmatrix}
    q^{\frac{1}{48}} \, \phi^*(q^{1/2})
    \\[2mm]
    q^{\frac{1}{3}} \, \nu^*(q^{1/2})
    \\[2mm]
    q^{\frac{1}{48}} \, \phi^*(-q^{1/2})
  \end{pmatrix} .
\end{equation}
Due to ~\eqref{phi_Eichler} --~\eqref{phi_Eichler_minus},
this is the Eichler integral 
of the vector-valued modular form with weight $3/2$;
\begin{equation}
  \boldsymbol{\Phi}_{2,3,4}(\tau)
  =
  \frac{1}{2} \sum_{n \in \mathbb{Z}}  \,
  n \,
  \begin{pmatrix}
    \chi_{24}^{(1)} (n)
    \\[2mm]
    \chi_{24}^{(2)} (n)
    \\[2mm]
    \psi_{6}^{(1)} (n)
  \end{pmatrix}
  \,
  q^{\frac{n^2}{48}} .
\end{equation}
The   modular $S$- and $T$-matrices under $\tau\to -1/\tau$ and
$\tau \to \tau+1$
are respectively  computed by use of~\eqref{Psi_under_S} and~\eqref{Psi_under_T} as
\begin{align}
  \label{S_T_234}
  \mathbf{S}
  & =
  \begin{pmatrix}
    1 & & \\[2mm]
    & 0 & 1 \\[2mm]
    & 1 & 0
  \end{pmatrix}
  ,
%%%%
  &
  \mathbf{T}
  & =
  \begin{pmatrix}
    & & \E^{\frac{1}{24} \pi \I}
    \\[2mm]
    & \E^{\frac{2}{3} \pi \I} &
    \\[2mm]
    \E^{\frac{1}{24} \pi \I} & & 
  \end{pmatrix} ,
\end{align}
which shows that the modular form
$[ \boldsymbol{\Phi}_{2,3,4}(\tau) ]^2$ is on $\Gamma(4)$, and
that  it has
the octahedral symmetry.
As a consequence the modular form $\boldsymbol{\Phi}_{2,3,5}(\tau)$
can be rewritten in terms of the Jacobi theta
functions~\cite{KHikami05a}.

We then obtain  an explicit form of the Eichler
integral~\eqref{Eichler_234} in
a limit  $\tau \to 1/N$   in terms of the Bernoulli polynomial as
 \eqref{Eichler_at_over_N}, and in
this case
it
gives the WRT invariant for the Seifert manifold
$M(2,3,4)$~\cite{KHikami05a}, whose fundamental group represents the
octahedral group;
\begin{theorem}
  \begin{equation}
    \E^{\frac{3 \pi \I}{2 N} } \,
    \left(
      \E^{\frac{2 \pi \I}{N}} - 1
    \right) \,
    \tau_N \left(
      M(2,3,4)
    \right)
    =
    \frac{\sqrt{2}}{4} \,
    \left( 1 + (-1)^N \right) \,
    \left(
      2 - 
      \phi^*(\E^{\frac{\pi \I}{N}})
    \right) .
  \end{equation}
\end{theorem}

By definition~\eqref{3_phi_star} of $\phi^*(q)$,
we can conclude that
the WRT invariant is a limiting value from outside the unit circle
of the Ramanujan mock theta
function of the third order.

Concerning  the function $\omega^*(q)$, 
we see from  \eqref{omega_star_Eichler}
that
 the Eichler integral~\eqref{define_Eichler} with $P=3$
is written as
\begin{equation}
  \widetilde{
    \boldsymbol{\Phi}}_{P=3}(\tau)
  =
  \frac{1}{2} \, q^{\frac{1}{12}} \,
  \begin{pmatrix}
    \omega^*( q^{1/4}) +  \omega^*(- q^{1/4})
    \\[2mm]
    \omega^*( q^{1/4}) -  \omega^*(- q^{1/4})
  \end{pmatrix} ,
\end{equation}
and a result of Ref.~\citen{KHikami05a} indicates the following;
\begin{theorem}
  \begin{equation}
    \E^{\frac{\pi \I}{2 N}} \,
    \left(
      \E^{\frac{2 \pi \I}{N}} - 1
    \right) \,
    \tau_N
    \left(
      M(2,2,3)
    \right)
    =
    1 - \omega^*(\E^{\frac{\pi \I}{2 N}})
    +
    \E^{\frac{N}{2} \pi \I} \,
    \left(
      1 - \omega^*(- \E^{\frac{\pi \I}{2 N}})
    \right)  .
  \end{equation}
\end{theorem}

To rewrite the quantum invariant in terms of the $q$-product,
it makes sense to give
the mock (false) theta functions $\phi^*(q)$ and $\nu^*(q)$ 
in the form which terminates at the finite order in the case of $q$
being root of unity.
\begin{prop}
  \begin{gather}
    \label{phi_star_finite}
    \phi^*(q)
    =
    1+
    q \sum_{n=0}^\infty
    (q; -q)_n \, q^n ,
    \\[2mm]
    \label{nu_star_finite}
    \nu^*(q)
    =
    \omega^*(q^2)
    =
    \sum_{n=0}^\infty
    (q^2; q^4)_n \, q^{2n} .
  \end{gather}
\end{prop}

\begin{proof}
  When we define the $q$-hypergeometric function by
  \begin{equation}
    F
    \left(
      \begin{matrix}
        a \\
        b
      \end{matrix}   \,
      ; q  , t
    \right)
    =
    \sum_{n=0}^\infty
    \frac{
      (a \, q)_n}{
      (b \, q)_n} \,
    t^n ,
  \end{equation}
  we have
  (see  (20.71) and (20.72) in Ref.~\citen{NJFine88Book})
  \begin{gather}
    \frac{1}{1+a} \,
    F
    \left(
      \begin{matrix}
        0 \\
        -a
      \end{matrix} \,
      ;
      q,a
    \right)
    =
    F
    \left(
      \begin{matrix}
        q^{-1} \\
        0
      \end{matrix} \,
      ;
      q^2, a^2
    \right) ,
    \\[2mm]
    (1-a) \,
    F
    \left(
      \begin{matrix}
        -1 \\
        0
      \end{matrix} \,
      ; q, a
    \right)
    =
    F
    \left(
      \begin{matrix}
        0 \\
        a
      \end{matrix} \,
      ; q^2,      a \,q 
    \right) .
  \end{gather}
  Applying these identities, we obtain the statement.
\end{proof}

Alternative computation of $q$-series
follows from topological fact that
the Seifert manifold $M(2,3,4)$ is constructed from $(+2)$-surgery on
the right-handed trefoil~\cite{LMose71a}.
Applying a surgery formula~\cite{LCJeff92a,ResheTurae91a}  to the
colored Jones polynomial~\eqref{our_Jones} for the trefoil, we get
\begin{equation}
  \phi^*(q)
  =
  1 + q
  \sum_{k \geq n \geq 0}^\infty
  (-1)^n \,
  \begin{bmatrix}
    k \\
    n
  \end{bmatrix}_{q^2} \,
  q^{n (2n+3) + k^2} .
\end{equation}

To close this section,
we note that \eqref{phi_star_finite} and~\eqref{nu_star_finite}
may give formulae as follows  as  a power series in $t$;
\begin{gather}
  \begin{aligned}
    \E^{-t/24} \,
    &
    \left(
      1+
      \sum_{n=0}^\infty
      \E^{-(n+1)t} \,
      (1 - \E^{-t}) \,
      (1 + \E^{-2t}) 
      \cdots
      (1 +(-1)^n \E^{-n t}) 
    \right)
    \\
    & \qquad \qquad
    =
    \sum_{k=0}^\infty
    \frac{
      L \left( -2 \, k , \chi_{24}^{(1)} \right) }{
      k!
    } \,
    \left(
      -\frac{t}{24}
    \right)^k ,
  \end{aligned}
  \\[2mm]
    \E^{-t/3}
    \sum_{n=0}^\infty
    \E^{-n t} \,
    (1 - \E^{-t}) \,
    (1 - \E^{-3 t}) \cdots
    (1 - \E^{-(2n-1)t})
    =
    \sum_{k=0}^\infty
    \frac{
      L \left( -2 \, k , \chi_{24}^{(2)} \right)
    }{
      k!
    } \,
    \left( - \frac{t}{48} \right)^k ,
  \end{gather}
  where
  \begin{gather*}
    2 \,
    \frac{\cos(3 \, x) \, \cos(2 \,x)}{
      \cos(6 \, x)
    }
    =
    \sum_{k=0}^\infty
    \frac{
      L \left( -2 \, k , \chi_{24}^{(1)} \right)
    }{
      \left(2 \, k \right) !
    } \,
    (-1)^k \,
    x^{2 k} ,
    \\[2mm]
    \frac{
      \cos(x) }{
      \cos(3 \, x)
    }
    =
    \sum_{k=0}^\infty
    \frac{
      L \left( -2 \, k , \chi_{24}^{(2)} \right)
    }{
      (2 \, k)!
    } \,
    (-1)^k \,
    \left(
      \frac{x}{2}
    \right)^{ 2 k} .
  \end{gather*}

%%%%%%%%%%%%%%%%
\section{The 7th Order Mock Theta Functions}
\label{sec:7th}

We continue to study the seventh order mock theta
functions~\cite{GEAndre86c,Hicker88b}.
There are 3 Ramanujan's mock theta functions, and they are read as
\begin{align}
  \mathcal{F}_0(q)
  & =
  \sum_{n=0}^\infty \frac{q^{n^2}}{(q^{n+1})_n} ,
  \\[2mm]
  \mathcal{F}_1(q)
  & =
  \sum_{n=1}^\infty \frac{q^{n^2}}{(q^{n})_n}
  ,
  \\[2mm]
  \mathcal{F}_2(q)
  & =
  \sum_{n=0}^\infty \frac{q^{n(n+1)}}{(q^{n+1})_{n+1}} .
\end{align}
These $q$-series have also meaning even when
$|q|>1$, and for our convention we define
\begin{align}
  \label{def_F0star}
  \mathcal{F}_0^*(q)
  & =
  \mathcal{F}_0(1/q)
  =
  \sum_{n=0}^\infty
  (-1)^n \,
  \frac{
    q^{\frac{1}{2} n(n+1)}
  }{
    (q^{n+1})_n
  } ,
  \\[2mm]
  \mathcal{F}_1^*(q)
  & =
  \mathcal{F}_1(1/q)
  =
  \sum_{n=1}^\infty
  (-1)^n \,
  \frac{q^{\frac{1}{2} n(n-1)}
  }{
    (q^{n})_n
  } ,
  \\[2mm]
  \mathcal{F}_2^*(q)
  & =
  q^{-1} \, \mathcal{F}_2(1/q)
  =
  -\sum_{n=0}^\infty
  (-1)^n \,
  \frac{
    q^{\frac{1}{2} n (n+3)}
  }{
    (q^{n+1})_{n+1}
  } .
\end{align}

By applying the method of the Bailey chain~\eqref{simple_Bailey}
with the Bailey pair  A(3), A(2), and A(4) in Slater's list~\cite{LJSlater51},
we can prove  that
these are the \emph{false} theta function in the sense of
Rogers~\cite{LJRogers17a}.
\begin{prop}
  \begin{gather}
    \mathcal{F}_0^*(q)
    =
    \sum_{n=0}^\infty
    \chi_{84}^{(1,1,1)}(n) \, q^{\frac{n^2 - 1}{168}} ,
    \\[2mm]
    \mathcal{F}_1^*(q)
    =
    \sum_{n=0}^\infty
    \chi_{84}^{(1,1,2)}(n) \, q^{\frac{n^2 - 25}{168}} ,
    \\[2mm]
    \mathcal{F}_2^*(q)
    =
    \sum_{n=0}^\infty
    \chi_{84}^{(1,1,3)}(n) \, q^{\frac{n^2 - 121}{168}} ,
  \end{gather}
  where $\chi_{84}^{\vv{\ell}}(n)$ is an odd periodic function with
  modulus $84$ defined by
  \begin{equation*}
    \begin{aligned}
      \chi_{84}^{(1,1,1)}(n)
      & =
      \psi_{84}^{(1)}(n) -       \psi_{84}^{(13)}(n)
      -    \psi_{84}^{(29)}(n) +       \psi_{84}^{(41)}(n) ,
      \\[2mm]
      \chi_{84}^{(1,1,2)}(n)
      & =
      - \psi_{84}^{(5)}(n) -   \psi_{84}^{(19)}(n)
      -   \psi_{84}^{(23)}(n) -     \psi_{84}^{(37)}(n) ,
      \\[2mm]
      \chi_{84}^{(1,1,3)}(n)
      & =
      - \psi_{84}^{(11)}(n) -    \psi_{84}^{(17)}(n)
      -    \psi_{84}^{(25)}(n) -     \psi_{84}^{(31)}(n) .
    \end{aligned}
  \end{equation*}
\end{prop}

% \begin{proof}
%   We apply a method of Bailey pair,
%   especially
%   A(3), A(2), and A(4) in Ref.~\citen{LJSlater51}
%   for these cases.
%   See Ref.~\citen{LJRogers17a}.
%%%%%%%%%%%%
%   For the first, we use
%   \begin{align}
%      \beta_n
%      & =
%      \frac{q^{n}}{(q)_{2 n}}
%      &
%      \begin{cases}
%        \alpha_{3k-1 > 0} = - q^{2 k (3 k-1)}
%        \\[2mm]
%        \alpha_{3k > 0} =  q^{2 k (3 k-1)} +  q^{2 k (3 k+1)},
%        \qquad
%        \alpha_0 = 1
%        \\[2mm]
%        \alpha_{3k+1 > 0} = - q^{2 k (3 k+1)}
%      \end{cases}
%    \end{align}
%    with $x=1$ which is Slater's A(3)~\cite{LJSlater51}.
%    For the second identity, we use A(2); $x=q$ and
%    \begin{align}
%      \beta_n
%      & =
%      \frac{1}{(q^2)_{2 n}}
%      &
%      \begin{cases}
%        \alpha_{3k-1 > 0} =  q^{ k (6 k-1)}
%        \\[2mm]
%        \alpha_{3k \geq 0} =    q^{ k (6 k+1)}
%        \\[2mm]
%        \alpha_{3k+1 > 0} = -q^{(2 k +1)(3k +1)} -  q^{(2 k+1) (3 k+2)}
%      \end{cases}
%    \end{align}
%    The last identities can be proved when we use
%    \begin{align}
%      \beta_n
%      & =
%      \frac{q^{n}}{(q^2)_{2 n}}
%      &
%      \begin{cases}
%        \alpha_{3k-1>0} =  q^{2 k (3 k-2)}
%        \\[2mm]
%        \alpha_{3k >0 } =    q^{2 k (3 k+2)},
%        \qquad
%        \alpha_0=1
%        \\[2mm]
%        \alpha_{3k+1 >0} = - q^{2 (k+1) (3 k+1)} - q^{2 k (3 k+2)}
%      \end{cases}
%    \end{align}
%    with  $x=q$ which is A(4) in Ref.~\citen{LJSlater51}.   
%\end{proof}

This proposition enables us to define~\cite{KHikami04b}
\begin{equation}
  \widetilde{\boldsymbol{\Phi}}_{2,3,7}(\tau)
  =
  \begin{pmatrix}
    q^{1/168} \, \mathcal{F}_0^*(q)
    \\[2mm]
    q^{25/168} \, \mathcal{F}_1^*(q)
    \\[2mm]
    q^{121/168} \, \mathcal{F}_2^*(q)
  \end{pmatrix} ,
\end{equation}
which can be regarded as the Eichler integral
of the vector-valued modular form with weight $3/2$;
\begin{equation}
  \boldsymbol{\Phi}_{2,3,7}(\tau)
  =
  \frac{1}{2} \, \sum_{n \in \mathbb{Z}} n
  \begin{pmatrix}
    \chi_{84}^{(1,1,1)}(n)
    \\[2mm]
    \chi_{84}^{(1,1,2)}(n)
    \\[2mm]
    \chi_{84}^{(1,1,3)}(n)
  \end{pmatrix}
  \, q^{\frac{n^2}{168}} .
\end{equation}
The  modular $S$- and $T$-matrices under $\tau \to -1/\tau$ and
$\tau \to \tau + 1$ are respectively given by
\begin{equation}
  \label{S_T_237}
  \begin{gathered}
    \mathbf{S}
    =
    - \frac{2}{\sqrt{7}}\,
    \begin{pmatrix}
      \sin( \frac{\pi}{7} ) &       \sin (\frac{2 \, \pi}{7}  )
      &
      \sin(\frac{3 \, \pi}{7})
      \\[2mm]
      \sin (\frac{2 \,\pi}{7}) &    -   \sin( \frac{3 \, \pi}{7} ) &
      \sin(\frac{ \pi}{7})
      \\[2mm]
      \sin (\frac{3 \, \pi}{7}) &       \sin( \frac{\pi}{7} ) &
      -\sin(\frac{2 \, \pi}{7})
    \end{pmatrix} ,
    \\[2mm]
    \mathbf{T}
    =
    \begin{pmatrix}
      \E^{ \frac{1}{84} \pi \I} &&
      \\
      &
      \E^{ \frac{25}{84} \pi \I}
      \\
      &&
      \E^{ - \frac{47}{84} \pi \I}
    \end{pmatrix}  .
  \end{gathered}
\end{equation}

Result in Ref.~\citen{KHikami04b} thus proves that
the mock (false) theta
function
$\mathcal{F}_0^*(q)$
gives  the WRT invariant for the Brieskorn homology sphere
$\Sigma(2,3,7)$ in a limit that $\tau$ goes to the $N$-th root of unity;

\begin{theorem}
  \begin{equation}
    \E^{-\frac{1}{42 N} \pi \I} \,
    \left(
      \E^{\frac{2 \pi \I}{N}} - 1
    \right) \,
    \tau_N
    \left(
      \Sigma(2,3,7)
    \right)
    =
    \frac{1}{2} \,
    \mathcal{F}_0^*(\E^{\frac{2\pi \I}{N}}) .
  \end{equation}
\end{theorem}

An explicit form of the right hand side  follows from
 \eqref{Eichler_at_over_N}.
Correspondingly an exact asymptotic expansion of the WRT invariant in
$N\to\infty$ can be computed  from  the nearly modular
property~\eqref{nearly_modular_Psi_tilde}~\cite{KHikami04b}.

By the   correspondence with the quantum invariant which is
originally  defined    for $q$ being root of unity,
it is tempting to
get the $q$-series expression which terminates at the finite order
when $q$ is  root of unity.
We obtain the following from  \eqref{def_F0star};
\begin{equation}
  \mathcal{F}_0^*(q) =
%   1 -
%   \sum_{n \geq m \geq 0}^\infty
%   (-1)^m \,
%   \begin{bmatrix}
%     n \\
%     m
%   \end{bmatrix}_q
%   q^{n+\frac{1}{2} m(m-1) + (n+m+1)^2}
  1 -
  q
  \sum_{k \geq n \geq 0}^\infty
  (-1)^n \,
  \begin{bmatrix}
    k    \\
    n
  \end{bmatrix}_q
  \,
  q^{(n+2) k - \frac{1}{2} n (n+1)} .
\end{equation}

Another construction of $q$-series follows from topological
understanding that the Brieskorn homology sphere $\Sigma(2,3,7)$ is
constructed by
$(-1)$-surgery on the right-handed trefoil~\cite{LMose71a}.
Applying a surgery formula  to the
colored Jones polynomial for the
trefoil~\eqref{our_Jones}
we obtain
\begin{equation}
  \mathcal{F}_0^*(q)
  =
  1-
  \sum_{k \geq n \geq 0}^\infty
  (-1)^n \,
  \begin{bmatrix}
    k \\
    n
  \end{bmatrix}_q \,
  q^{k+ \frac{1}{2} n(n-1) + (k+n+1)^2} .
\end{equation}
A
simpler expression was given in Ref.~\citen{TQLe03a} where
used was the
cyclotomic expansion of the  colored Jones
polynomial for trefoil~\eqref{cyclotomic_Jones}.

%%%%%%%%%%%%%%
\section{The 6th Order Mock Theta Functions}
\label{sec:6th}

There are 7 sixth order mock theta functions in 
Ramanujan's ``lost''
notebook~\cite{Ramanujan87Book},
and
we treat 3 functions among them defined by
\begin{gather}
  \varphi(q)
  =
  \sum_{n=0}^\infty
  (-1)^n \, q^{n^2} \,
  \frac{
    (q; q^2)_n
  }{
    (-q)_{2 n}
  } ,
  \\[2mm]
  \psi(q)
  =
  \sum_{n=0}^\infty
  (-1)^n \, q^{(n+1)^2} \,
  \frac{(q; q^2)_n}{
    (-q ; q)_{2 n +1}
  } ,
  \\[2mm]
  \rho(q)
  =
  \sum_{n=0}^\infty
  q^{n(n+1)/2} \,
  \frac{
    (-q)_n
  }{
    (q; q^2)_{n+1}
  } .
\end{gather}
Other functions can be written in terms of these functions and theta
functions as was proved in Ref.~\citen{AndreHicke91a}.
These $q$-series converge also for $|q|>1$, and
for our convention we define
\begin{gather}
  \varphi^*(q)
  =
  \varphi(1/q)
  =
  \sum_{n=0}^\infty
  q^n \,
  \frac{
    (q ; q^2)_n
  }{
    (-q)_{2 n}
  } ,
  \\[2mm]
  \psi^*(q)
  =
  \psi(1/q)
  =
  \sum_{n=0}^\infty
  q^n \,
  \frac{
    (q; q^2)_n
  }{(-q)_{2n+1}}  ,
  \\[2mm]
  \rho^*(q)
  =
  -q^{-1} \, \rho(1/q)
  =
  \sum_{n=0}^\infty
  (-1)^n \, q^n \,
  \frac{
    (-q)_n
  }{
    (q; q^2)_{n+1}
  }
  .
\end{gather}

These new functions are also the false theta functions
\emph{\`{a} la} Rogers as follows; 
\begin{prop}
  \begin{gather}
    \label{varphi_star}
    \varphi^*(q)
    =
    \sum_{n=0}^\infty
    \left(
      \psi_{12}^{(1)}(n) + \psi_{12}^{(5)}(n)
    \right)
    \, q^{\frac{n^2 -1}{24}} ,
    \\[2mm]
    \label{psi_star}
    \psi^*(q)
    =
    \sum_{n=0}^\infty
    \psi_{12}^{(3)}(n)   \, q^{\frac{n^2 - 9}{24}} ,
    \\[2mm]
    \label{rho_star}
    \rho^*(q)
    =
    \psi^*(q^2)
    =
    \sum_{n=0}^\infty
    \psi_{24}^{(6)}(n) \,
    q^{\frac{n^2 - 36}{48}}  ,
  \end{gather}
  where the odd periodic function $\psi_{2P}^{(a)}(n)$ is defined in
   \eqref{periodic_psi}.
%   where
%   \begin{align*}
%     \chi_{12}^{(1)}(n)
%     & = \psi_{12}^{(1)}(n) + \psi_{12}^{(5)}(n)
%     \\[2mm]
% %
%     \chi_{12}^{(2)}(n)
%     & = \psi_{12}^{(3)}(n)
%   \end{align*}
\end{prop}
\begin{proof}
  We do not have a simple proof. Anyway
  we recall the transformation formula of the $q$-hypergeometric functions as
  (see, \emph{e.g.}, Ref.~\citen{GEAndre81b})
  \begin{equation}
    \label{Andrews_formula}
    \sum_{n=0}^\infty
    \frac{
      (\alpha ;q^2)_n \,
      (\beta)_{2n}
    }{
      (q^2; q^2)_n \,
      (\gamma)_{2n}
    } \,
    z^n
    =
    \frac{
      (\beta)_\infty \, (\alpha \, z ; q^2)_\infty
    }{
      (\gamma)_\infty \, (z ; q^2)_\infty
    } \,
    \sum_{m=0}^\infty
    \frac{
      \left( \frac{\gamma}{\beta} \right)_m \,
      (z ; q^2)_m
    }{
      (q)_m \, (\alpha \, z ; q^2)_m
    } \,
    \beta^m  ,
  \end{equation}
  and another identity
  (see, \emph{e.g.},  (25.96) in Ref.~\citen{NJFine88Book})
  \begin{equation}
    \label{Fine_formula}
    \sum_{m=0}^\infty
    \frac{
      (\alpha \, q)_{2m} \, (\beta \, q)_m
    }{
      (\alpha  \, q)_m \, (q)_m
    } \, z^m
    =
    \frac{
      (\beta \, z \, q)_\infty
    }{
      (z)_\infty
    } \,
    \sum_{k=0}^\infty
    \frac{
      (\beta \, q)_k \, (z)_k
    }{
      (q)_k \, (\beta \, z \, q)_{2 k}
    } \,
    ( - \alpha \, z)^k \,
    q^{\frac{1}{2} k (3k+1)} .
  \end{equation}
  To prove  \eqref{psi_star}, we compute as follows;
  \begin{align*}
    \psi^*(q)
    & =
    \frac{1}{1+q} 
    \sum_{n=0}^\infty
    q^n \,
    \frac{(q)_{2n}}{
      (q^2; q^2)_n \, (-q^2)_{2n}
    }
    \\
    & =
%    \stackrel{\eqref{Andrews_formula}}{=}
    \frac{
      (q^2;q^2)_\infty
    }{
      (-q)_\infty
    } \,
    \sum_{m=0}^\infty
    q^m \,
    \frac{
      (-q)_m \, (q; q^2)_m
    }{
      (q)_m
    }
    \tag{by \eqref{Andrews_formula} with $\alpha=0$, $\beta=q$,
      $\gamma=-q^2$, $z=q$}
    \\
    & =
    (q)_\infty
    \sum_{m=0}^\infty
    q^m \,
    \frac{
      (q)_{2m}
    }{
      \left[ (q)_m \right]^2
    }
    \\
    & =
%    \stackrel{\eqref{Fine_formula}}{=}
    \sum_{k=0}^\infty
    (-q)^k \, q^{\frac{3}{2} k (k+1)} .
    \tag{by \eqref{Fine_formula} with $\alpha=1$, $\beta=0$, $z=q$}
  \end{align*}
  A proof of  \eqref{varphi_star} follows in the same manner;
  \begin{align*}
    \varphi^*(q)
    & =
    \sum_{n=0}^\infty
    q^n \, \frac{ (q)_{2n}}{
      (q^2; q^2)_n \, (-q)_{2n}
    }
    \\
    & =
% \stackrel{\eqref{Andrews_formula}}{=}
    \frac{(q)_\infty}{
      (-q)_\infty \, (q; q^2)_\infty
    }
    \sum_{m=0}^\infty
    q^m \,
    \frac{(-1)_m \, (q; q^2)_m}{(q)_m}
    \tag{by \eqref{Andrews_formula} with $\alpha=0$, $\beta=q$,
      $\gamma=-q$, $z=q$}
    \\
    & =
    (q)_\infty \,
    \left(
      1 +2 \, q \, \sum_{m=0}^\infty q^m \,
      \frac{(q^2)_{2m}}{
        (q)_m \, (q^2)_m
      }
    \right)
    \\
    & =
%\stackrel{\eqref{Fine_formula}}{=}
    \sum_{n \in \mathbb{Z}}
    (-1)^n \,
    q^{\frac{1}{2} n (3n-1)}
    + 2 \sum_{k=0}^\infty
    (-1)^k \,q^{\frac{1}{2} (k+1)(3k+2)} .
    \tag{by \eqref{triple_product} and \eqref{Fine_formula} with $\alpha=q$, $\beta=0$, $z=q$}
  \end{align*}

  Identity~\eqref{rho_star} can be proved as follows;
  \begin{align*}
    \rho^*(q)
    & =
    \frac{1}{1-q} \sum_{n=0}^\infty
    (-q)^n \,
    \frac{(q^2; q^2)_n}{
      (q)_n \, (q^3; q^2)_n
    }
    \\
    &  =
%\stackrel{\eqref{Andrews_formula}}{=}
    (q^2;q^2)_\infty
    \sum_{n=0}^\infty
    q^{2n} \,
    \frac{
      (q; q^2)_n \, (-q)_{2n}
    }{
      (q^2; q^2)_n
    }
    \tag{by \eqref{Andrews_formula} with $\alpha=q$, $\beta=-q$,
      $\gamma=0$, $z=q^2$}
    \\
    & =
    (q^2 ; q^2)_\infty
    \sum_{n=0}^\infty
    q^{2n} \,
    \frac{(-q^2 ; q^2)_n \,
      (q^2; q^4)_n
    }{
      (q^2; q^2)_n
    }
    \\
    & =
%\stackrel{\eqref{Andrews_formula}}{=}
    (-q^4; q^2)_\infty \, (q^2; q^4)_\infty
    \sum_{n=0}^\infty
    q^{2n} \,
    \frac{
      (q^2;q^2)_{2n}
    }{
      (q^4; q^4)_n \,
      (-q^4; q^2)_{2n}
    }
    \tag{by \eqref{Andrews_formula} with $q\to q^2$ and $\alpha=0$, $\beta=q^2$,
      $\gamma=-q^4$, $z=q^2$}
    \\
    & =
    \sum_{n=0}^\infty
    q^{2n} \,
    \frac{
      (q^2; q^4)_n
    }{
      (-q^2; q^2)_{2n+1}
    } ,
%     =
%     \psi^*(q^2)
  \end{align*}
  which completes the proof.
\end{proof}

Looking at ~\eqref{varphi_star} and~\eqref{psi_star}, we set the
vector-valued  functions as
\begin{equation}
  \boldsymbol{\Phi}_{2,3,3}(\tau)
  =
  \frac{1}{2} \,
  \sum_{n \in \mathbb{Z}}  \,
  n \,
  \begin{pmatrix}
    \frac{1}{\sqrt{2}} 
    \left(
      \psi_{12}^{(1)}(n) + \psi_{12}^{(5)}(n)
    \right)
    \\[2mm]
    \psi_{12}^{(3)}(n)
  \end{pmatrix}
  \,
  q^{\frac{n^2}{24}} .
\end{equation}
By the Poisson summation formula we find that this is a modular form with weight
$3/2$, and that
$S$- and $T$-matrices are respectively given by
\begin{align}
  \mathbf{S}
  & =
  \frac{1}{\sqrt{3}} \,
  \begin{pmatrix}
    1 & \sqrt{2}
    \\[2mm]
    \sqrt{2} & -1
  \end{pmatrix}
  ,
  &
  \mathbf{T}
  & =
  \begin{pmatrix}
    \E^{\frac{1}{12} \pi \I} &
    \\[2mm]
    & \E^{\frac{3}{4} \pi \I}
  \end{pmatrix}
  .
\end{align}
The modular form
$[ \eta(\tau) ]^{-1} \cdot
\boldsymbol{\Phi}_{2,3,3}(\tau)$
is on $\Gamma(3)$, which has the symmetry of the tetrahedral group.
By definition, the mock theta functions can be regarded as the Eichler
integral of this modular form;
\begin{equation}
  \widetilde{
    \boldsymbol{\Phi}}_{2,3,3} ( \tau)
  =
  \begin{pmatrix}
    q^{\frac{1}{24}} \, \varphi^*(q)
    \\[2mm]
    q^{\frac{3}{8}} \, \psi^*(q)
  \end{pmatrix} .
\end{equation}

The Eichler integral $\widetilde{\boldsymbol{\Phi}}_{2,3,3}(\tau)$
has appeared as the WRT invariant for the Seifert
manifold $M(2,3,3)$~\cite{KHikami05a} whose fundamental group is
the tetrahedral group;
\begin{theorem}
  \begin{multline}
    \E^{\frac{\pi \I}{N}} \,
    \left(
      \E^{\frac{2 \pi \I}{N}} - 1 
    \right) \, \tau_N \left(
      M(2,3,3)
    \right)
    \\
    =
    \frac{1+2 \, \E^{\frac{2}{3} \pi \I N}}{\sqrt{3}} \,
    \left(
      1 - \frac{1}{2} \, \varphi^*(\E^{\frac{2\pi \I}{N}})
    \right)
    -
    \frac{1 - \E^{\frac{2}{3} \pi \I N}}{\sqrt{3}} \,
    \E^{\frac{2 \pi \I}{3 N}} \,
    \psi^*(\E^{\frac{2 \pi \I}{N}})
    .
  \end{multline}
\end{theorem}

As was noticed in Ref.~\citen{KHikami05a}, the WRT invariants for the
manifold $M(2,2,6)$ and $M(2,3,3)$ are related to each other, and we
also  obtain the WRT invariant for $M(2,2,6)$ as
\begin{theorem}
  \begin{equation}
    \E^{\frac{2 \pi \I}{N}} \,
    \left(
      \E^{\frac{2 \pi \I}{N}} - 1
    \right) \,
    \tau_N\left(
      M(2,2,6)
    \right)
    =
    2 \,
    \left(
      1 - \varphi^*(\E^{\frac{2 \pi \I}{N}})
    \right)
  \end{equation}
\end{theorem}

The function
$q^{1/8} \, \rho^*(q^{1/6}) =
q^{1/8} \, \psi^*(q^{1/3})$ is also regarded as the Eichler
integral of the weight-$3/2$
modular form $[ \eta(\tau) ]^3$,
which, by the Jacobi triple product
formula~\eqref{triple_product},
is written in an infinite sum as
\begin{equation}
  \left[
    \eta(\tau)
  \right]^3
  =
  \sum_{n=0}^\infty
  (-1)^n \,
  ( 2 \, n + 1 ) \,
  q^{\frac{1}{8} (2 n +1)^2} .
\end{equation}
Recalling a result in Ref.~\citen{KHikami05a},
we  can conclude that it gives the WRT invariant for the Seifert manifold
$M(2,2,2)$.
\begin{theorem}
  \begin{equation}
    \left(
      \E^{\frac{2 \pi \I}{N}} - 1
    \right) \,
    \tau_N \left(
      M(2,2,2)
    \right)
    =
    2 \,
    \left(
      1 - 2 \, \rho^* ( \E^{\frac{\pi \I}{3 N}})
    \right) .
  \end{equation}
\end{theorem}

%%%%%%%%%%%%%%%
\section{The 10th Order Mock Theta Functions}
\label{sec:10th}

Remaining Ramanujan's mock theta functions are the  tenth order.
There are 4 functions in Ref.~\citen{Ramanujan87Book}
(see also Ref.~\citen{YSCho99a}), and they are defined by
\begin{gather}
  \Phi(q)
  =
  \sum_{n=0}^\infty 
  \frac{q^{n(n+1)/2}}{(q;q^2)_{n+1}}  ,
  \\[2mm]
  \Psi(q)
  =
  \sum_{n=0}^\infty
  \frac{q^{(n+1)(n+2)/2}}{(q;q^2)_{n+1}} ,
  \\[2mm]
  X(q)
  =
  \sum_{n=0}^\infty
  (-1)^n \,
  \frac{
    q^{n^2}}{
    (-q)_{2n}
  } ,
  \\[2mm]
  \chi(q)
  =
  \sum_{n=0}^\infty
  (-1)^n \,
  \frac{
    q^{(n+1)^2}}{
    (-q)_{2n+1}
  } .
\end{gather}
All these defining $q$-series  converge  also  for $|q|>1$, and as before
we define our functions as follows;
\begin{gather}
  \Phi^*(q)
  =
  - q^{-1} \, \Phi(1/q)
  =
  \sum_{n=0}^\infty (-1)^n \,
  \frac{q^{n(n+3)/2}}{
    (q;q^2)_{n+1}
  } ,
  \\[2mm]
  \Psi^*(q)
  =
  - \Psi(1/q)
  =
  \sum_{n=0}^\infty
  (-1)^n \,
  \frac{
    q^{n(n+1)/2}
  }{
    (q;q^2)_{n+1}
  } ,
  \\[2mm]
  X^*(q) = X(1/q)
  =
  \sum_{n=0}^\infty
  (-1)^n \,
  \frac{q^{n (n+1)}}{
    (-q)_{2n}
  }  ,
  \\[2mm]
  \chi^*(q) = \chi(1/q)
  =
  \sum_{n=0}^\infty
  (-1)^n \,
  \frac{q^{n (n+1)}}{
    (-q)_{2n+1}
  } .
\end{gather}

We have the following;
\begin{prop}
  \begin{gather}
    \Phi^*(q)
    =
    \sum_{n=0}^\infty
    \left(
      \psi_{10}^{(2)}(n)
      +
      \psi_{10}^{(3)}(n)
    \right) \,
    q^{\frac{1}{5}(n^2 - 4)} ,
    \\[2mm]
    \Psi^*(q)
    =
    \sum_{n=0}^\infty
    \left(
      \psi_{10}^{(1)}(n)
      +
      \psi_{10}^{(4)}(n)
    \right) \,
    q^{\frac{1}{5}(n^2 - 1)} ,
    \\[2mm]
    X^*(q)
    =
    \sum_{n=0}^\infty
    \psi_{10}^{(1)}(n)
    q^{\frac{1}{40}(n^2 - 1)} ,
    \\[2mm]
    \chi^*(q)
    =
    \sum_{n=0}^\infty
    \psi_{10}^{(3)}(n)
    q^{\frac{1}{40}(n^2 - 9)} .
  \end{gather}
\end{prop}
\begin{proof}
  All these identities follow from  \eqref{simple_Bailey} with the
  Bailey pair,
  C(4), C(3), G(3), and  G(2) in Slater's list~\cite{LJSlater51}
  respectively.
\end{proof}

Our functions are regarded as the Eichler integral~\eqref{define_Eichler}
of the modular
form~\eqref{define_modular_Psi} with $P=5$.
A result in Ref.~\citen{KHikami05a} proves  that these give
the WRT invariant for the Seifert manifold
$M(2,2,5)$.
\begin{theorem}
  \begin{multline}
    \E^{\frac{3 \pi \I}{2 N}} \,
    \left(
      \E^{\frac{2 \pi \I}{N}} - 1
    \right) \,
    \tau_N
    \left(
      M(2,2,5)
    \right)
    \\
    =
    1 + \E^{-\frac{N}{2} \pi \I} -
    \left(
      1- \E^{- \frac{N}{2} \pi \I}
    \right) \,
    \Psi^*(\E^{\frac{\pi \I}{2 N}})
    - 2 \,
    \E^{-\frac{N}{2} \pi \I} \,
    X^*(\E^{\frac{4 \pi \I}{N}}) .
  \end{multline}
\end{theorem}

By applying  \eqref{q-binomial_over} to the defining $q$-series,
it is straightforward to get the infinite $q$-series,
which terminates at
the finite order for root of unity,
of the quantum  invariants.

%%%%%%%%%%%%%%%
\section{Discussions}

We have shown that the Ramanujan mock theta functions are related to
the quantum invariant for the  Seifert manifolds $M(p_1,p_2,p_3)$.
We see that
some of the mock theta functions can be defined also even when $q$ is
outside the unit circle, and that,
by replacing $q$ by $1/q$,
they become the \emph{false} theta functions in a sense of Rogers.
This substitution $q \to 1/q$ seems to correspond to study,
not directly the Eichler integral
$\widetilde{\Psi}_P^{(a)}(\tau)$~\eqref{define_Eichler},
but
the Eichler
integral $\widehat{\Psi}_P^{(a)}(z)$~\eqref{hat_Eichler} defined in the
lower half plane,
which has a nice transformation property~\eqref{modular_Psi_hat}.
These \emph{false} theta functions have already appeared in studies of
the WRT invariant for the Seifert manifold as the Eichler integral of
the half-integral weight modular form.
Namely a limiting value of the Eichler integral in $\tau\to 1/N$ for
$N\in\mathbb{Z}_{>0}$ gives the WRT invariant $\tau_N(\mathcal{M})$.
Combining these results, we   can deduce a remarkable connection
between the Ramanujan mock theta functions and the quantum invariants.
We summarize this correspondence in Table~\ref{tab:mock_order}.
\begin{table}[htbp]
  \centering
  \renewcommand{\arraystretch}{1.4}
  \begin{tabular}{cc|c}
    order & mock theta functions  & 3-manifolds $\mathcal{M}$
    \\
    \hline \hline
    2  ?
    & $D_5(q)$ ?
    &
    $M(2,2,2)$
    \\
    3  & $\phi(q)$ & $M(2,3,4)$
    \\
    & $\omega(q)$ &  $M(2,2,3)$
    \\
    4  ?
    & $D_6(q)$ ?
    &
    $M(2,2,4)$
    \\
    5  & $\chi_0(q)$ & Poincar{\'e} sphere $\Sigma(2,3,5)$
    \\
    6  & $\varphi(q)$, $\psi(q)$,
    [$\rho(q)$]
    & $M(2,3,3)$,  $M(2,2,6)$,
    [$M(2,2,2)$]
    \\
    7  & $\mathcal{F}_0(q)$ & Brieskorn sphere $\Sigma(2,3,7)$
    \\
    8 ?
    & $I_{12}(q)$ ?
    &
    $M(2,2,8)$
    \\
    10  & $\Psi(q)$, $X(q)$ & $M(2,2,5)$
    \\
    \hline
  \end{tabular}
  \caption{Mock theta functions as the SU(2) WRT invariant
    $\tau_N(\mathcal{M})$ for the
    Seifert manifolds
    $\mathcal{M}=M(p_1,p_2,p_3)$.
    Functions of order 2, 4, and 8 are our proposal based on the WRT
    invariants.
  }
  \label{tab:mock_order}
\end{table}
Unfortunately we do not find such correspondence for the eighth order
mock theta functions proposed in Ref.~\citen{GordMcIn00a}.
Motivated from Ref.~\citen{LJRogers17a,LJSlater51},
we may  expect that the functions defined by
\begin{gather}
  D_5(q)
  =
  \sum_{n=0}^\infty
  q^n \,
  \frac{
    (-q)_n
  }{
    (q; q^2)_{n+1}
  } ,
  \\[2mm]
  D_6(q)
  =
  \sum_{n=0}^\infty
  q^n \,
  \frac{
    (-q^2; q^2)_n
  }{
    (q^{n+1})_{n+1}
  } ,
  \\[2mm]
  I_{12}(q)
  =
  \sum_{n=0}^\infty
  q^{2n} \,
  \frac{
    (-q; q^2)_n
  }{
    (q^{n+1})_{n+1}
  } ,
  \\[2mm]
  I_{13}(q)
  =
  \sum_{n=0}^\infty
  q^{n} \,
  \frac{
    (-q; q^2)_n
  }{
    (q^{n+1})_{n+1}
  } ,
\end{gather}
will be related to the 2nd/4th/8th  order mock theta functions.
We  can introduce the $q$-series from above definitions by replacing
$q$ with $1/q$ as
\begin{gather}
  D_5^*(q)
  =
  -q^{-1} \, D_5(1/q)
  =
  \sum_{n=0}^\infty
  (-1)^n q^{n(n+1)/2} \,
  \frac{
    (-q)_n
  }{
    (q; q^2)_{n+1}
  } ,
  \\[2mm]
  D_6^*(q)
  =
  -q^{-1} \, D_6(1/q)
  =
  \sum_{n=0}^\infty
  (-1)^n
  q^{n (n+1)/2} \,
  \frac{
    (-q^2; q^2)_n
  }{
    (q^{n+1})_{n+1}
  }  ,
  \\[2mm]
  I_{12}^*(q)
  =- q^{-1} \, I_{12}(1/q)
  =
  \sum_{n=0}^\infty
  (-1)^n \,
  q^{n (n+1)/2} \,
  \frac{
    (-q;q^2)_{n}
  }{
    (q^{n+1})_{n+1}
  } ,
  \\[2mm]
  I_{13}^*(q)
  =- q^{-1} \, I_{13}(1/q)
  =
  \sum_{n=0}^\infty
  (-1)^n \,
  q^{n (n+3)/2} \,
  \frac{
    (-q;q^2)_{n}
  }{
    (q^{n+1})_{n+1}
  } ,
\end{gather}
and
the Bailey pairs, D(5), D(6), I(12), and I(13), in Slater's list
prove the following;
\begin{prop}
  \begin{gather}
    D_5^*(q)
    =
    \sum_{n=0}^\infty
    \psi_4^{(1)}(n) \, q^{\frac{1}{4} ( n^2 - 1)}  ,
    \\[2mm]
    D_6^*(q)
    =
    \sum_{n=0}^\infty
    \left(
      \psi_8^{(1)}(n)
      +
      \psi_8^{(3)}(n)
    \right) \,
    q^{\frac{1}{4} ( n^2 - 1)} ,
    \\[2mm]
    I_{12}^*(q)
    =
    \sum_{n=0}^\infty
    \left(
      \psi_{16}^{(1)}(n) +       \psi_{16}^{(7)}(n)
    \right) \,
    q^{\frac{1}{16}(n^2 - 1)} ,
    \\[2mm]
    I_{13}^*(q)
    =
    \sum_{n=0}^\infty
    \left(
      \psi_{16}^{(3)}(n) +       \psi_{16}^{(5)}(n)
    \right) \,
    q^{\frac{1}{16}(n^2 - 9)} .
  \end{gather}
\end{prop}
This proves
\begin{equation*}
  \rho^*(q) = \psi^*(q^2) = D_5^*(q^3)
\end{equation*}
where $\rho^*(q)$ and $\psi^*(q)$ are from the 6-th
order~\eqref{psi_star},~\eqref{rho_star}.
{}From computations in Ref.~\citen{KHikami05a},
these give the WRT invariants as follows;
\begin{theorem}
  \begin{gather}
    \left(
      \E^{\frac{2 \pi \I}{N}} - 1
    \right) \,
    \tau_N
    \left(
      M(2,2,2)
    \right)
    =
    2 \,
    \left(
      1
      - 2 \,
      D_5^*(\E^{\frac{\pi \I}{N}})
    \right)
    \\[2mm]
    \left(
      \E^{\frac{2 \pi \I}{N}} - 1
    \right) \,
    \tau_N
    \left(
      M(2,2,4)
    \right)
    =
    \left(
      1 + (-1)^N
    \right) \,
    \left(
      1 - D_6^*( \E^{ \frac{\pi \I}{2 N}} )
    \right) ,
    \\[2mm]
    \E^{\frac{3 \pi \I}{2 N}} \,
    \left(
      \E^{\frac{2 \pi \I}{N}} - 1
    \right) \,
    \tau_N
    \left(
      M(2,2,8)
    \right)
    =
    \left(
      1 + (-1)^N
    \right) \,
    \left(
      1 - I_{12}^*( \E^{ \frac{\pi \I}{ N}} )
    \right) .
  \end{gather}
\end{theorem}

As the WRT invariant for wide class of the Seifert manifolds
has a nearly modular
property~\cite{KHikami04b,KHikami04e,KHikami05a,KHikami04f},
we may derive  mock  (false) theta functions from explicit form of
these invariants.
In such process, the colored Jones polynomial for torus
knots~\cite{KHikami04a}
and the twist knots~\cite{GMasb03a} would be helpful.

%%%%%%%%%%%
\section*{Acknowledgments}
The author would like to thank Thang Le for explaining his
construction of the quantum invariants.
He also thanks T.~Takata for useful discussions.
This work is supported in part by the Grant-in-Aid for Young
Scientists
from the Ministry of Education, Culture, Sports, Science and
Technology of Japan.
%%%%%%%%%%%%%%
%\newpage
%\bibliographystyle{physics}
% %\bibliographystyle{amsalpha}
% %\bibliographystyle{JHEP}
% %\bibliographystyle{siam}
\bibliographystyle{alphaKH}
% %ibliographystyle{klunum}
%%%%%%%%%%%%%%%%%%
%\bibliography{_def,gravity,square,math,ba,tba,math5,vm,square2,math4,qalg,math3,math2,poisson,geometry,soliton,cft,knot,tqft,comb,number}

\end{document}